\documentclass[12pt]{article}
\usepackage[xdvi]{graphicx}
\usepackage{amssymb}
\newcommand{\bea}{\begin{eqnarray}}
\newcommand{\eea}{\end{eqnarray}}

\makeatletter
\@addtoreset{equation}{section}
\makeatother

\def\ignore#1{{}}

\begin{document}
\begin{titlepage}
\begin{flushright}
OU-HET 699/2011
\end{flushright}

\vspace{30ex}

\begin{center}
{\Large\bf
Most attractive channel
for self-breaking

\vspace{1.5ex}
of grand unification group symmetry
}
\end{center}

\vspace{5ex}

\begin{center}
{\large
Nobuhiro Uekusa
}
\end{center}
\begin{center}
{\it Department of Physics, 
Osaka University \\
Toyonaka, Osaka 560-0043
Japan} \\
\textit{E-mail}: uekusa@het.phys.sci.osaka-u.ac.jp
\end{center}


\vspace{8ex}

\begin{abstract}

Most attractive channel 
is studied for SU(5) grand unified theory
in five dimensions where
{\bf 5*} and {\bf 10} fermions and
a {\bf 24} gauge boson propagate in
the bulk.
If there are bulk fermions
with zero mode for the chirality opposite
to quark and lepton multiplets,
they can contribute to
forming {\bf 5}, {\bf 24}, {\bf 50} and {\bf 75}
scalar bound states.
We find that the inclusion of 
a simple anomaly-free set for
zero mode
yields scalar bound states
whose binding strengths are
the largest for
{\bf 24} and the second largest
for {\bf 5}.
This corresponds to the breaking pattern of the
SU(5) to color and electromagnetism
via the standard model gauge group.

\end{abstract}

\end{titlepage}


\newpage

\section{Introduction}

While elementary particles have
mass and charge,
the origin still remains unknown.
When the generation of masses 
is related to symmetry breaking,
the corresponding problem is how symmetry is broken.
In the standard model, the Higgs mechanism
describes physics of elementary 
particles very well up to the weak scale.
However, it has been found in the own framework
of the standard model that 
this picture is not valid at higher
energy scales.
In addition, the standard model does not have 
any explanation for quantization of charge.
Hence, 
we need a deeper understanding for mass and charge,
or symmetry breaking and charge
beyond the standard model.

If extra dimensions are included in an effective
theory,
the strength of force would behave 
unlike four dimensions. 
The gauge bosons of the standard model
propagating in extra dimensions
can rapidly become strongly coupled and
form scalar bound states of quarks and leptons.
The self-breaking of the standard-model
gauge symmetry was proposed in Ref.~\cite{ArkaniHamed:2000hv}. 
The authors proposed that
the existence of a Higgs doublet is a consequence of 
the standard-model gauge symmetry and three generation
of quarks and leptons
provided the gauge bosons and fermions
propagate in appropriate extra dimensions compactified
at a TeV scale.
It has been shown earlier
that electroweak symmetry may be broken
by fields propagating in extra dimensions
in Ref.~\cite{Dobrescu:1998dg}\cite{Cheng:1999bg}.
Also in the Randall-Sundrum warped space 
\cite{Randall:1999ee}\cite{Randall:1999vf},
transition of the strength of force
and the type of bound states have been studied
\cite{Davoudiasl:1999tf}-%
\cite{Uekusa:2011ud}.

As for quantization of charge,
grand unification
has attracted much attention.
The charges belong to subgroups of a unification group.
Quantum numbers for quarks and leptons
are fixed by the group structure.
The SU(3), SU(2) and U(1) couplings meet at a single
point and  the couplings
are evaded from blowup of 
the values by renormalization group evolution.
Beyond the unification scale,
the energy dependence of 
the three couplings are shown 
graphically as one identical line 
due to contributions of twelve $X$ and $Y$ bosons.
Thus, a grand unification with extra dimensions
might contain the generation of masses
and the quantization of charges automatically.

In this paper,
we study most attractive channel
for an SU(5) grand unification in five dimensions.
Our setup is that components in 
each SU(5) multiplet obey
the same boundary condition with respect 
to extra dimensions
although gauge symmetry breaking
by boundary conditions might
be an interesting possibility
\cite{Kawamura:2000ev}
\cite{Kawamura:2000ir}.
The energy dependence 
of couplings in gauge theories 
with extra dimensions
has been found 
\cite{Dienes:1998vh}
\cite{Dienes:1998vg}
and has been examined in detail
\cite{Bhattacharyya:2006ym}
\cite{Uekusa:2007im}.
In particular,
 the energy dependence of 
the three couplings should be shown as
one identical line
beyond the unification scale.
It has been emphasized in
Ref.~\cite{Uekusa:2008iz} that
this is not the case
for gauge symmetry breaking by boundary conditions
where the same boundary conditions do not span
irreducible representations of the unification group.
In order to keep gauge coupling 
unification above the unification scale,
we consider 
only the self-breaking for the 
origin of symmetry breaking.
The fundamental fields are 
a gauge boson and fermions.
A gauge boson belongs to {\bf 24}.
Quarks and leptons are included in
{\bf 5*} and {\bf 10} Dirac fermions
whose left-handed components have
zero mode.
If these fields are only objects
propagating in the bulk,
there are no 
scalar bound states composed of zero mode.
When an anomaly-free set of fields with 
zero mode for the right-handed component
is added,
we find scalar bound states
whose binding strengths are
the largest for
{\bf 24} and the second largest
for {\bf 5},
correspondingly to
the breaking pattern
SU(5) $\to$ SU(3)$\times$SU(2)$\times$U(1)
$\to$ SU(3)$\times$U(1).

The paper is organized as follows.
In Section~\ref{memo},
we start with the basics such as
the scenario, 
framework, field content and group structure.
With consideration for anomaly-free sets,
scalar bound states and the binding strengths 
are given
in Section~\ref{sec:mini}.
Here we find various composites.
The simple model to form {\bf 24} and {\bf 5}
for appropriate symmetry breaking
is proposed.
Furthermore, the composites {\bf 50} and {\bf 75}
are discussed for the doublet-triplet splitting.
We conclude in Section~\ref{concl}.

\section{Basics of the model \label{memo}}

We consider a gauge theory
with two branes on fixed points
of the orbifold $S^1/Z_2$ in five dimensions.
Bulk fields can have strong coupling compared with
the corresponding four-dimensional theory
so that the ordinary four-dimensional massless gluon
and quarks
does not give rise to gauge symmetry breaking
 and that the five-dimensional effect
may change 
potentials for composites.
If
non-vanishing vacuum expectation values
are generated,
symmetry is broken.
The pattern of symmetry breaking depends
on the attractive force of composites
and the masses of the constituents.
The analysis can be applied 
for both of flat spacetime and warped spacetime.

In the present model, there are no fundamental scalar
fields. 
The gauge boson $A_M$ has 
the four-dimensional vector component $A_\mu$
and the four-dimensional scalar component $A_5$.
The boundary conditions at the location
of the branes are given
by Neumann for all $A_\mu$ and
Dirichlet for all $A_5$.
Quarks and leptons are included 
as zero mode for {\bf 5*}
and {\bf 10} Dirac fermions, 
$\psi_{5*}$ and $\psi_{10}$.
For the left-handed component, 
the Neumann boundary condition is imposed.
Since singlet neutrinos do not produce attractive force,
they can be omitted in the present analysis.
For these boundary conditions,
fields which belong to
irreducible representations of SU(5)
obey the same boundary conditions
so that at higher energy scales 
above the scale of
symmetry breaking by any condensation
the theory behave as SU(5).
For simplicity, the number of the generation is 
chosen as one.

The source of symmetry breaking 
is scalar bound states represented by 
the form $\bar{\psi}_I\psi_J$ where
$I,J$ are denoted as the species of fermions.
The attractive force is dominantly originated 
from exchange of a gauge boson between fermions
and is determined by the gauge coupling and the group structure.
As for the group structure, we take the notation
$\textrm{tr}(t_r^a t_r^b) = C(r)
\delta^{ab}$
where $t_r^a$ is denoted as
the representation matrices in the irreducible
representation $r$
and $C(r)$ is a constant for each representation $r$.
The relation
$d(r) C_2(r) = d(\textrm{Adj}) C(r)$
is fulfilled where
$C_2(r)$ is the quadratic Casimir operator
 for each representation and
 $d$ is the dimension of each representation.
The values $C$, $C_2$ and $d$ are summarized 
for several irreducible representations
of SU($N$)
in Table~\ref{tab:sun}.
For $N=5$,
the diagonal generator
$t^8 = 
   \textrm{diag}(2, 2, 2, -3, -3 )
  /(2\sqrt{15})$
plays the same role as the U(1) hypercharge
up to a overall factor.
We call its charge $Q_Y$.
For SU(3)$\times$SU(2)$\times$U(1),
 $\psi_{5*}$ and $\psi_{10}$
are decomposed into
({\bf 3*}, {\bf 1})${}_{Q_Y=-\sqrt{15}/15}$
$\oplus$
({\bf 1}, {\bf 2})${}_{Q_Y=\sqrt{15}/10}$ 
and
({\bf 3}, {\bf 2})${}_{Q_Y=-\sqrt{15}/30}$
 $\oplus$
({\bf 3*}, {\bf 1})${}_{Q_Y=2\sqrt{15}/15}$
 $\oplus$
({\bf 1}, {\bf 1})${}_{Q_Y=-\sqrt{15}/5}$, respectively.
\begin{table}[htb]
\caption{SU($N$) property. Each box in Young tableaux
is
symmetric for the horizontal direction
and antisymmetric for the vertical direction.
\label{tab:sun}}
\begin{center}
\begin{tabular}{cc|ccc}
\hline \hline 
&& $d$ & $C$ & $C_2$ \\
\hline
\framebox{} 
&& $N$ & ${1\over 2}$ & ${N^2-1\over 2N}$ \\
\begin{picture}(5,15)(0,0)
\put(0,6.5){\framebox{}}
\put(0,0){\framebox{}}
\end{picture}
&&
 ${N(N-1)\over 2}$ 
 & ${N-2\over 2}$
 & ${(N+1)(N-2)\over N}$ 
\\
\begin{picture}(10,15)(0,0)
\put(0,0){\framebox{}}
\put(6.5,0){\framebox{}}
\end{picture}
&&
   ${N(N+1)\over 2}$ 
 & ${N+2\over 2}$
 & ${(N-1)(N+2)\over N}$ 
\\ 
\begin{picture}(5,25)(0,0)
\put(0,0){\framebox{}}
\put(0,6.5){\framebox{}}
\put(0,13){\framebox{}}
\end{picture}
&&
   ${N(N-1)(N-2)\over 6}$ 
 & ${(N-2)(N-3)\over 4}$
 & ${3(N-3)(N+1)\over 2N}$ 
\\ 
\begin{picture}(10,25)(0,0)
\put(0,0){\framebox{}}
\put(0,6.5){\framebox{}}
\put(6.5,6.5){\framebox{}}
\end{picture}
&&
   ${N(N^2-1)\over 6}$ 
 & ${(N^2-3)\over 2}$
 & ${3(N^2-3)\over 2N}$ 
 \\ 
\begin{picture}(15,15)(0,0)
\put(0,0){\framebox{}}
\put(6.5,0){\framebox{}}
\put(13,0){\framebox{}}
\end{picture}
&&
   ${N(N+1)(N+2)\over 6}$ 
 & ${(N+2)(N+3)\over 4}$
 & ${3(N-1)(N+3)\over 2N}$ 
\\ 
 \begin{picture}(5,15)(0,0)
 \put(0,0){Adj}
\end{picture}
&&
   $N^2-1$ 
 & $N$
 & $N$ 
\\ 
\hline
\end{tabular}
\end{center}
\end{table}

For the quark and lepton multiplets,
possible combinations of the bound states are
$\bar{\psi}_{5*}\psi_{5*}$,
$\bar{\psi}_{10}\psi_{10}$ and
$\bar{\psi}_{5*}\psi_{10}$.
They are written in terms of four-dimensional fields
dependent on the five-dimensional coordinates
as chirality mixing operators
$\bar{\psi}_{L}\psi_{R} +
\bar{\psi}_{R}\psi_{L}$.
Because $\psi_R$ consists of Kaluza-Klein massive
mode without zero mode,
it is difficult to have potentials 
for bound states
to yield nonzero vacuum expectation values.
Hence, we need to add new fields with
zero mode for the right-handed component.
In order that $\psi_{5*}$ and $\psi_{10}$ do not
form brane mass terms together with the new fields,
candidates are except for {\bf 5*} and {\bf 10}.
A singlet is also excluded
because it does not contribute to attractive force.

For forming the fundamental {\bf 5} or 
the anti-fundamental {\bf 5*} 
scalar bound states together with
quark and lepton multiplets,
possible bulk fermions are {\bf 5}, 
{\bf 10*} and {\bf 24}.
For forming an adjoint {\bf 24} scalar bound state
together with quark and lepton multiplets,
possible bulk fermions are {\bf 15}, 
{\bf 40*} and {\bf 45}.
Added new fermions have zero mode only for
the right-handed component.
In general they are chiral in four dimensions.
Therefore new fields need to be added in
such a way that anomaly is canceled.
An anomaly coefficient $A(r)$ is defined by
\bea
   \textrm{tr}\left[t^a_r\{t^b_r,t^c_r\}\right]
   = {1\over 2} A(r) d^{abc}
\eea
where
$\{t_N^a, t_N^b\}
  ={1\over N} \delta^{ab} + d^{abc} t_N^c$
for SU($N$).
For $N=5$, the generators associated with 
$Q_Y$ yield $d^{888}=-1/\sqrt{15}$.
The anomaly coefficients for the above irreducible
representations are given by
\bea
   A({\bf 5}) = -{1\over 2},~
   A({\bf 10*}) = {1\over 2},~ 
   A({\bf 15}) = {9\over 2}, ~
   A({\bf 24}) = 0, ~
   A({\bf 40*}) = -8 ,~
   A({\bf 45}) = 3 .
\eea
Thus the minimal anomaly-free set 
to obtain {\bf 5} or {\bf 5*} scalar bound states
and a {\bf 24} scalar bound state
is ({\bf 10*}, {\bf 15}, 
{\bf 40*}, {\bf 45}) Dirac fermions.
In the next section, we will examine the binding
strengths for bound states in the most attractive 
channel approximation.

\section{Composites and binding strengths
\label{sec:mini}}

In this section 
we derive
possible bound states 
and their quantum numbers and binding strengths
by adding bulk fields
with zero mode for the right-handed component, $\chi$.
Here $\bar{\psi} \chi$ includes 
$\bar{\psi}_L^{(0)} \chi_R^{(0)}$  
composed of only zero mode
and  strong attractive force
can lead to potentials with phase transition.
First we examine
the minimal anomaly-free set.
Taking into account the pattern of
symmetry breaking more,
we study a simple anomaly-free set.
In addition, the {\bf 50} and {\bf 75}
bound states associated with the 
doublet-triplet splitting are discussed.

\subsubsection*{The minimal anomaly-free set}

The quark and lepton multiplets are
$\psi_{5*}$ and $\psi_{10}$.
The minimal set of
addition fermions is given by
$\chi_{10*}$, $\chi_{15}$,
$\chi_{40*}$ and $\chi_{45}$. 
It is necessary to examine
the binding strengths for eight possible scalar
bound states
\bea
   \bar{\psi}_{5*} \chi_{10*} , ~
   \bar{\psi}_{5*} \chi_{15} , ~
   \bar{\psi}_{5*} \chi_{40*} , ~
   \bar{\psi}_{5*} \chi_{45} ,
 ~
   \bar{\psi}_{10} \chi_{10*} ,~
   \bar{\psi}_{10} \chi_{15} , ~
   \bar{\psi}_{10} \chi_{40*} ,~
   \bar{\psi}_{10} \chi_{45} .
    \label{psichi}
\eea
They are decomposed further in terms of
irreducible representations into twenty four
 bound states     
\bea
   {\bf 5} \otimes {\bf 10*}
  &\!\!\!=\!\!\!&
   {\bf 5*} \oplus {\bf 45} ,
\\
   {\bf 5} \otimes {\bf 15}  
 &\!\!\!=\!\!\!&
   {\bf 35} \oplus {\bf 40} ,  \label{5-15}
\\
    {\bf 5} \otimes {\bf 40*}
  &\!\!\!=\!\!\!&
   {\bf 10*} \oplus {\bf 15*} \oplus 
   {\bf 175^*_\textrm{\scriptsize A}} ,
\\
   {\bf 5} \otimes {\bf 45}
  &\!\!\!=\!\!\!&
   {\bf 24} \oplus {\bf 75} \oplus {\bf 126} ,
\\
  {\bf 10*} \otimes {\bf 10*}
  &\!\!\! =\!\!\!& 
    {\bf 5} \oplus {\bf 45*} \oplus {\bf 50*} ,
\\
 {\bf 10*} \otimes {\bf 15}
    &\!\!\!=\!\!\!&
       {\bf 24} \oplus {\bf 126} ,  \label{10-15}
\\
  {\bf 10*} \otimes {\bf 40*}
     &\!\!\!=\!\!\!&
       {\bf 24}  \oplus {\bf 75}
       \oplus {\bf 126*} \oplus 
  {\bf 175^*_{\textrm{\scriptsize B}}},
\\
  {\bf 10*} \otimes {\bf 45}
   &\!\!\!=\!\!\!&
    {\bf 10} \oplus {\bf 15}\oplus
     {\bf 40*} \oplus {\bf 175} \oplus {\bf 210} ,
\eea 
The binding strength for $\bar{\psi}\chi$
is given by \cite{Raby:1979my}\cite{ArkaniHamed:2000hv}
\bea
  {1\over 2}g^2 \left[ C_2(\bar{\psi})
    + C_2 (\chi) - C_2(\bar{\psi}\chi) \right] .
\eea
The coupling constant $g$ is common 
for types of fields in the present
model.  
For SU(5), the values 
of $C_2$ for the irreducible representations
 appearing here are summarized in Table~\ref{tab:c2}.
\begin{table}[htb]
\caption{The quadratic Casimir operator $C_2$
for each representation $r$.
\label{tab:c2}}
\begin{center}
\vspace{-2ex}
\begin{tabular}{c|ccccc ccccc ccc}
\hline\hline
$r$ & 
{\bf 5} & 
{\bf 10} & 
{\bf 15} & 
{\bf 24} & 
{\bf 35} & 
{\bf 40} & 
{\bf 45} &
{\bf 50} &  
{\bf 75} & 
{\bf 126} & 
{\bf 175${}_{\textrm{\scriptsize A}}$} & 
{\bf 175${}_{\textrm{\scriptsize B}}$} &
{\bf 210} 
\\ \hline
$C_2$ &
${12\over 5}$  & 
${18\over 5}$  &
${28\over 5}$  &
5  & 
${48\over 5}$  & 
${33\over 5}$  & 
${32\over 5}$  & 
${42\over 5}$  & 
8  &
10 & 
${48\over 5}$  & 
12 & 
${90\over 7}$ \\
\hline
\end{tabular}
\end{center}
\end{table}
Using these values, 
we find the binding strengths for the twenty four
bound states as shown
in Table~\ref{tab:bind}.
\begin{table}[htb]
\caption{Binding strengths for $\bar{\psi}\chi$
for the minimal set in unit of $g^2$.
\label{tab:bind}}
\vspace{-2ex}
\begin{center}
\begin{tabular}{ccc}
 \hline\hline
  Composite &
  Constituents &
  Binding \\
 &&
strength 
  \\
  \hline
{\bf 10} &
$\bar{\psi}_{10} \chi_{45}$ &
$16/5$
\\
{\bf 10*} &
$\bar{\psi}_{5*} \chi_{40*}$ &
$27/10$
\\
{\bf 24} &
$\bar{\psi}_{10} \chi_{40*}$ &
$13/5$
\\
{\bf 5} &
$\bar{\psi}_{10} \chi_{10*}$ &
$12/5$
\\
{\bf 15} &
$\bar{\psi}_{10} \chi_{45}$ &
$11/5$
\\
{\bf 24} &
$\bar{\psi}_{10} \chi_{15}$ &
$21/10$
\\
{\bf 24} &
$\bar{\psi}_{5*} \chi_{45}$ &
$19/10$
\\
{\bf 5*} &
$\bar{\psi}_{5*} \chi_{10*}$ &
$9/5$
\\
{\bf 15*} &
$\bar{\psi}_{5*} \chi_{40*}$ &
$17/10$
\\
{\bf 40*} &
$\bar{\psi}_{10} \chi_{45}$ &
$17/10$
\\
{\bf 75} &
$\bar{\psi}_{10} \chi_{40*}$ &
$11/10$
\\
{\bf 40} &
$\bar{\psi}_{5*} \chi_{15}$ &
$7/10$
\\ \hline
\end{tabular}
~
\begin{tabular}{ccc}
 \hline\hline
  Composite &
  Constituents &
  Binding \\
 &&
strength 
  \\
  \hline
{\bf 75} &
$\bar{\psi}_{5*} \chi_{45}$ &
$2/5$
\\
{\bf 45*} &
$\bar{\psi}_{10} \chi_{10*}$ &
$2/5$
\\
{\bf 175${}_{\textrm{\scriptsize A}}$}&
$\bar{\psi}_{10} \chi_{45}$ &
$1/5$
\\
{\bf 126*} &
$\bar{\psi}_{10} \chi_{40*}$ &
$1/10$
\\
{\bf 45} &
$\bar{\psi}_{5*} \chi_{10*}$ &
$-1/5$
\\
{\bf 175${}_{\textrm{\scriptsize A}}$*} &
$\bar{\psi}_{5*} \chi_{40*}$ &
$-3/10$
\\
{\bf 126} &
$\bar{\psi}_{10} \chi_{15}$ &
$-2/5$
\\
{\bf 126} &
$\bar{\psi}_{5*} \chi_{45}$ &
$-3/5$
\\
{\bf 50*} &
$\bar{\psi}_{10} \chi_{10*}$ &
$-3/5$
\\
{\bf 35} &
$\bar{\psi}_{5*} \chi_{15}$ &
$-4/5$
\\
{\bf 175${}_{\textrm{\scriptsize B}}$*} &
$\bar{\psi}_{10} \chi_{40*}$ &
$-9/10$
\\
{\bf 210} &
$\bar{\psi}_{10} \chi_{45}$ &
$-10/7$
\\ \hline
\end{tabular}
\end{center}
\end{table}

For the minimal field content, it is found that
the binding strengths for {\bf 10} and {\bf 10*} are
larger than that of the adjoint {\bf 24}.
Usually SU(5) symmetry is broken by
{\bf 24} or {\bf 75}.
The results given in
Table~\ref{tab:bind} means
SU(5) symmetry would not be broken desirably
unless there is some mechanism such that 
the binding for {\bf 10} and {\bf 10*}
 decreases effectively 
or the binding for {\bf 24} increases.
In the following
we consider a non-minimal but simple anomaly-free set
where this problem does not arise.

\subsubsection*{The simple anomaly-free set}

As given in the previous section,
a {\bf 24} scalar bound state can be made of 
{\bf 15}, {\bf 40*} and {\bf 45} Dirac fermions 
with quark and lepton multiplets.
From Table~\ref{tab:bind}, it is read that
a composite {\bf 24} has the largest binding strength
 only for a {\bf 15} $\chi$ field and
the other composites are largest for different
quantum numbers:
the composite {\bf 10*} for 
a {\bf 40*} $\chi$ field and 
the composite {\bf 10} for 
a {\bf 45} $\chi$ field.
Therefore the fermion with zero mode for 
the right-handed component
to form a {\bf 24} scalar bound state is uniquely 
fixed as a {\bf 15} $\chi$ field.

One of the simplest anomaly-free fermion content
with {\bf 15}
is 9 $\chi_5$ and 1 $\chi_{15}$.
The composite are given by
$\bar{\psi}_{5*}\chi_5$ 
$\bar{\psi}_{5*}\chi_{15}$
$\bar{\psi}_{10}\chi_5$
and $\bar{\psi}_{10}\chi_{15}$
as well as 8 copies of $\chi_5$.
They are decomposed into eight 
irreducible representations,   
\bea
   {\bf 5} \otimes 
   {\bf 5} &\!\!\!=\!\!\!& 
  {\bf 10}  \oplus {\bf 15} ,
\\
  {\bf 10*} \otimes {\bf 5} &\!\!\!=\!\!\!&
  {\bf 5*} \oplus {\bf 45} ,
\eea
with Eqs.~(\ref{5-15}) and (\ref{10-15}).
For this simple set, the binding strengths are
given in Table~\ref{tab:binds}.
\begin{table}[htb]
\caption{Binding strengths for $\bar{\psi}\chi$
for the simple set in unit of $g^2$.
\label{tab:binds}}
\vspace{-2ex}
\begin{center}
\begin{tabular}{ccc}
 \hline\hline
  Composite &
  Constituents &
  Binding \\
 &&
strength 
  \\
  \hline
{\bf 24} &
$\bar{\psi}_{10} \chi_{15}$ &
$21/10$
\\
{\bf 5*} &
$\bar{\psi}_{10} \chi_{5}$ &
$9/5$
\\
{\bf 40} &
$\bar{\psi}_{5*} \chi_{15}$ &
$7/10$
\\
{\bf 10} &
$\bar{\psi}_{5*} \chi_{5}$ &
$3/5$
\\ \hline
\end{tabular}
~
\begin{tabular}{ccc}
 \hline\hline
  Composite &
  Constituents &
  Binding \\
 &&
strength 
  \\
  \hline
{\bf 45} &
$\bar{\psi}_{10} \chi_{5}$ &
$-1/5$
\\
{\bf 126} &
$\bar{\psi}_{10} \chi_{15}$ &
$-2/5$
\\
{\bf 15} &
$\bar{\psi}_{5*} \chi_{5}$ &
$-2/5$
\\
{\bf 35} &
$\bar{\psi}_{5*} \chi_{15}$ &
$-4/5$
\\ \hline
\end{tabular}
\end{center}
\end{table}

We find scalar bound states
whose binding strengths are
the largest for
{\bf 24} in $\bar{\psi}_{10} \chi_{15}$ 
 and the second largest
for {\bf 5} in
$\bar{\chi}_{5}\psi_{10}$.
The largest strength for {\bf 24}
corresponds to the breaking of the
SU(5) to SU(3)$\times$SU(2)$\times$U(1).
After the SU(5) is broken,
nine {\bf 5} and {\bf 5*} composite scalar 
fields contribute
for the breaking of
SU(2)$\times$U(1) to U(1).

Now we compare 
the binding strengths given in Table~\ref{tab:binds}
with the results of the standard model gauge group.
When quarks $Q,U,D$, leptons $L,E,N$
and SU(3)$\times$SU(2)$\times$U(1) gauge bosons
propagate in five dimensions,
the binding strengths for the composites
with zero mode are given in Table~\ref{tab:bs}.
Here $Q,L$ are SU(2) doublets and
$U,D,E,N$ are SU(2) singlets.
\begin{table}[htb]
\begin{center}
\caption{Binding strength for the standard model gauge 
group \cite{Uekusa:2011ud}.
\label{tab:bs}}
\begin{tabular}{c|cccc}
\hline\hline
 Composite
& Constituents
& SU(3)$\times$ SU(2)$\times$U(1)
& Binding
& Binding for
\\
&
& representation
& strength
& $\sqrt{3\over 5}\,g_1 = g_2=g_3\equiv g$ 
\\
\hline
$H_1$ & $\bar{Q}U$ 
& $({\bf 1}, {\bf 2},{1\over 2})$
& ${4\over 3} g_3^2 + {1\over 9} g_1^2$
& (7/5) $g^2$
\\
$H_2$ & $\bar{Q}D$
& $({\bf 1}, {\bf 2},-{1\over 2})$
& ${4\over 3} g_3^2 -{1\over 18} g_1^2$
& (13/10) $g^2$ 
\\ \hline\hline
$S_{10}$ & $\bar{E}Q$
& $({\bf 3}, {\bf 2},{7\over 6})$
& $-{1\over 6}g_1^2$
& ($-1/10$) $g^2$
\\
$S_{11}$ & $\bar{L}U$
& $({\bf 3}, {\bf 2},{5\over 6})$
& 0 & 0
\\
$S_{12}$ & $\bar{L}D$
& $({\bf 3}, {\bf 2},{1\over 6})$
& ${1\over 6}g_1^2$
& (1/10) $g^2$
\\
$S_{13}$ & $\bar{L}E$
& $({\bf 1}, {\bf 2},-{1\over 2})$
& ${1\over 2} g_1^2$
& (3/10) $g^2$
\\
$S_{14}$
& \multicolumn{2}{l}{
$N$ or $\bar{N}$ is included}
 & 0 & 0
\\
\hline
\end{tabular}
\end{center}
\end{table}

From Tables~\ref{tab:binds} and \ref{tab:bs},
the binding strengths for {\bf 24} and {\bf 5*}
are larger than that of the Higgs doublet $H_1$.
Because the self-breaking of the standard model
gauge group has been claimed
\cite{ArkaniHamed:2000hv}
\cite{Dobrescu:1998dg}
\cite{Cheng:1999bg},
the symmetry breaking associated
with the larger binding strengths
can occur. 
In other words, the symmetry breaking of
SU(5) to color and electromagnetism can be expected.

\subsubsection*{On the composites 50, 50* and 75}

A solution to the doublet-triplet splitting
has been to break the
SU(5) symmetry by the real representation
{\bf 75} instead of {\bf 24} 
and add {\bf 50} and {\bf 50*}  
\cite{Masiero:1982fe}
\cite{Grinstein:1982um}.
Now we consider the composites
{\bf 50}, {\bf 50*} and {\bf 75}.
      
First we examine the composite {\bf 75}.
For quark-lepton multiplets      
$\psi_{5*}$ and $\psi_{10}$, 
fermions $\chi$
with the representations {\bf 10}, {\bf 40}, 
{\bf 45} and {\bf 50}
can form {\bf 75} bound states as
\bea
  {\bf 5} \otimes {\bf 45} 
&\!\!\!=\!\!\!& {\bf 24} \oplus {\bf 75}  \oplus {\bf 126},
  \label{45}
\\
  {\bf 5} \otimes {\bf 50}
 &\!\!\!=\!\!\!&
  {\bf 75} \oplus {\bf 175_{\textrm{\bf\scriptsize B}}} ,
\\
  {\bf 10*} \otimes {\bf 10}
 &\!\!\!=\!\!\!&
  {\bf 1} \oplus {\bf 24} \oplus {\bf 75} ,
\\
  {\bf 10*} \otimes {\bf 40*}
  &\!\!\!=\!\!\!&
    {\bf 24} \oplus {\bf 75} \oplus
   {\bf 126*} \oplus {\bf 175_{\textrm{\bf \scriptsize B}}*} .
\eea
Among these constituents,
only the {\bf 50} $\chi$ field
 yields a {\bf 75} bound state
without the adjoint {\bf 24}. 
The binding strengths
for the constituents $\bar{\psi}_{5*}\chi_{50}$   
are 12/5 for the composite {\bf 75}
and $-3/5$ for the composite 
{\bf 175${}_{\textrm{\scriptsize B}}$}.
This point is favorable.
However,
in addition to the coupling with $\psi_{5*}$,
$\bar{\psi}_{10}\chi_{50}$ 
needs to be taken into account
where
${\bf 10*} \otimes
    {\bf 50} = {\bf 10} \oplus {\bf 175} 
  \oplus {\bf 315}$.
The binding strength
for the constituents $\bar{\psi}_{10}\chi_{50}$   
are 21/5 for the composite {\bf 10}.
This means that the composite {\bf 75}
with the quark-lepton multiplets for the constituents 
involves other composites bound 
with larger strengths.

Next we examine the composites {\bf 50} and {\bf 50*}.
When the {\bf 5*} and {\bf 10} quark-lepton multiplets 
and their conjugates coupled to 
fermions with right-handed zero mode 
form {\bf 50} or {\bf 50*} bound states,
the composites would include
{\bf 45} or {\bf 45*} bound states simultaneously.
Since the quadratic Casimir operators 
have the relation $C_2({\bf 50}) < C_2({\bf 45})$,
the composite {\bf 45} is bound stronger than
the composite {\bf 50}.
This gives rise to SU(2) doublet mass terms.
It is because the representation {\bf 45} for SU(5)
has a color singlet and weak doublet state
for SU(3)$\times$SU(2)
and the composites
{\bf 45} and {\bf 75} can become an SU(5) singlet with 
the composite {\bf 5} as seen in Eq.~(\ref{45}).

Therefore
doublet-triplet splitting by
the composites {\bf 50}, {\bf 50*} and {\bf 75}
would not occur minimally.
The model would need to be modified for
application of the idea.

\section{Conclusion \label{concl}}

If extra dimensions are included in an effective
theory,
the strength of force would behave 
unlike four dimensions. 
We have studied most attractive channel
for an SU(5) grand unification in five dimensions.
We have assumed that components in 
each SU(5) multiplet obey
the same boundary condition with respect 
to extra dimensions to keep the gauge coupling
unification above the unification scale.
In addition to quarks, leptons and gauge bosons,
$\psi_{5*}$, $\psi_{10}$ and $A_M$,
anomaly-free sets of fields $\chi$ with 
zero mode for the right-handed component
lead to various scalar bound states.

The minimal anomaly-free set 
to obtain {\bf 5} and {\bf 24} scalar bound states
is ($\psi_{5*}$, $\psi_{10}$, $A_M$)
plus ($\chi_{10*}$, $\chi_{15}$, 
$\chi_{40*}$, $\chi_{45}$).
For this field content, the
binding strengths for {\bf 10} and {\bf 10*} have been found to be 
larger than that of {\bf 24}.
Then, before SU(5) symmetry is broken to 
the standard model gauge group by {\bf 24},
symmetry breaking can occur by {\bf 10} and
{\bf 10*}.

A simple way to overcome this problem
is to adopt ($\chi_5^{I=1,\cdots,9}$, $\chi_{15}$)
instead of the anomaly-free set
($\chi_{10*}$, $\chi_{15}$, 
$\chi_{40*}$, $\chi_{45}$).
Here the {\bf 15} $\chi$ field is uniquely chosen
because it is the only representation
to form a {\bf 24} scalar bound state 
whose binding strength is large
compared with other composites.  
The largest strength for the composite {\bf 24}
corresponds to the breaking of the
SU(5) to the standard model gauge group.
After the SU(5) is broken,
{\bf 5} in $\bar{\psi}_{10}\chi_5$ contribute
for the breaking of
SU(2)$\times$U(1) to U(1).
In addition, we have found that
the binding strengths for ${\bf 24}$ and ${\bf 5}$
are larger than the known results for 
the composite Higgs doublet
for the standard model gauge group.

We have shown that the symmetry breaking of
the grand unification group 
to color and electromagnetism may occur
by {\bf 24} and {\bf 5}.
It would be more interesting 
to identify other fermion sets to assure
the double-triplet splitting.
However, we have found that
the doublet-triplet splitting by
the composites {\bf 50}, {\bf 50*} and {\bf 75}
does not seem to occur minimally.

Finally,
our analysis can be developed in 
various directions such as modification of
 field contents.
It needs to be examined in more detail
how
masses and charges of
elementary
particles are derived appropriately.

\vspace{4ex}
\subsubsection*{Acknowledgments}

This work is supported by Scientific Grants 
from the Ministry of Education
and Science, Grant No.~20244028.





\vspace*{10mm}


\end{document}